\documentstyle[onecolumn]{mn}
\input epsf
\newcommand{\Fs}{\,^*\!\bmath{F}}
\newcommand{\FF}{\bF\cdot^*\!\!\!\bmath{F}}
\newcommand{\bF}{\bmath{F}}
\newcommand{\bT}{\bmath{T}}

\newcommand{\bu}{\bmath{u}}
\newcommand{\bU}{\bmath{U}}
\newcommand{\bV}{\bmath{V}}
\newcommand{\bQ}{\bmath{Q}}
\newcommand{\bP}{\bmath{P}}
  
\title{On the Properties of Time-Dependent, Force-Free, Degenerate Electrodynamics} 
\author[S.S.Komissarov] 
{ Komissarov S.S. \\  
Department of Applied Mathematics, 
The University of Leeds, Leeds.   
E-mail: serguei@amsta.leeds.ac.uk }

\begin{document} 
\label{firstpage} 
\maketitle 

\begin{abstract} 
This   paper   formulates   time-dependent,   force-free,   degenerate
electrodynamics as  a hyperbolic system of conservations  laws.  It is
shown  that this  system  has  four characteristic  modes,  a pair  of
fast waves propagating with the speed of light and a pair of Alfv\'en 
waves.  All these modes are linearly degenerate. 
The results of this analytic  study can be used in developing upwind
numerical  schemes for the  electrodynamics of  black hole  and pulsar
magnetospheres. As an example, this paper describes a simple one-dimensional 
numerical scheme based on linear and exact Riemann solvers. 
\end{abstract} 

\section{Introduction} 

In magnetospheres of pulsars and black holes the electromagnetic field
is  so strong  that the  inertia  and pressure  of the  plasma can  be
ignored.   As a result 
the Lorentz  force, $F_{\alpha\beta}J^\alpha$, vanishes and
the   transport   of   energy   and  momentum   is entirely
electromagnetic  \cite{GJ69,BZ77}.   This  justifies the  name
``force-free'' to  describe the  electrodynamics of pulsars  and black
holes.  However,  the electrodynamics of the  magnetospheres is rather
different from  that in a  vacuum which is obviously  also force-free.
Indeed,  the  magnetospheric plasma  is  plentiful  enough to  support
strong electric  currents and screen  the electric field
\cite{GJ69,BZ77}.  An electromagnetic  field satisfying this condition
is  called ``degenerate'' \cite{MT82}.   Thus, the  electrodynamics of
pulsar  and  black   hole  magnetospheres  is  force-free,  degenerate
electrodynamics (FFDE). 

So far,  theoretical studies of  pulsar and black  hole magnetospheres
have  assumed  a steady-state,  indeed  the  properties of FFDE 
as  a system of time-dependent equation  have not been
studied   systematically.    It   is   only   recently   that   Uchida
\shortcite{Uchida}  has  developed a  rather  elegant  theory of  this
system in which the electromagnetic field is described in terms of two
scalar   functions,  called   ``Euler  potentials''.    However,  this
formulation  is  not  particularly  suitable  for  numerical  analysis
because its basic equations, when written in components, involve mixed
space and  time second order  derivatives.  In this paper,  we present
another formulation in which the  evolution equations have the form of
hyperbolic   conservation   laws.    We   also   describe   a   simple
one-dimensional,  upwind  numerical  scheme  based on  these  analytic
results.  This scheme can easily be generalized to multidimensions and
curved space-time.  In  fact, we have already constructed  a 2D scheme
for  the  Kerr  space-time  and  used it  \cite{Kom01}  to  study  the
Blandford-Znajek   mechanism  \cite{BZ77}   for   the  extraction   of
rotational energy from  black holes.

\section{Evolution Equations of Degenerate Electrodynamics} 

Since far away from the relativistic ``star'' the inertia of particles is bound 
to become 
important, the approximation of ideal relativistic magnetohydrodynamics (RMHD) 
could be considered as more satisfactory in general \cite{Phinney}. 
This suggests one way of deriving the time-dependent 
equations of FFDE, namely by considering the low inertia limit of RMDH. 
We shell adopt this approach because it reveals the 
close connection between these two systems. However,
the time-dependent equations of FFDE
can also be derived entirely within the frame of electrodynamics in a rather obvious
way. Besides that, some of 
the microphysical conditions of MHD, like the small mean free path of plasma 
particles, do not have to be satisfied in FFDE.

In the covariant form the equations of RMHD are as follows  
\cite{Lichner,Anile} 

\begin{description} 

\item{Maxwell's equation:}
\begin{equation} 
  \nabla_\alpha  \Fs^{\alpha \beta} = 0, 
\label{FE}
\end{equation}

\item{Energy-momentum equation:}
\begin{equation} 
  \nabla_\alpha \bT^{\alpha \beta} = 0, 
\label{EME}
\end{equation}

\item{Continuity equations}
\begin{equation} 
  \nabla_\alpha (n_{(i)} \bu^\alpha ) = 0 ,
\label{CE}
\end{equation} 

\end{description} 
where $\bT^{\alpha\beta}$ is the total stress-energy-momentum tensor, 
$\Fs^{\alpha  \beta}$ is  the  dual tensor  of electromagnetic  field,
$n_{(i)}$ is the proper number  density for species $i$ and $\bu^\alpha$
is the fluid 4-velocity. The total stress-energy-momentum tensor

\begin{equation} 
   \bT^{\alpha\beta} = \bT^{\alpha\beta}_{(m)} + 
                     \bT^{\alpha\beta}_{(f)},  
\end{equation} 
is the sum of the stress-energy-momentum tensor of matter,   
$\bT^{\alpha \beta}_{(m)}$, 
and the stress-energy-momentum tensor of electromagnetic field 

\begin{equation} 
   \bT^{\alpha\beta}_{(f)} = \bF^{\alpha}_{\ \mu}  
             \bF^{\beta\mu} -\frac{1}{4}
             (\bF_{\mu\nu}\bF^{\mu\nu}) \bmath{g}^{\alpha\beta}, 
\label{Temf}
\end{equation} 
where $\bmath{g}_{\alpha\beta}$ is the  metric tensor and 
$\bF_{\alpha\beta}$ is
the electromagnetic field tensor. In this paper we assume that 
Greek indices run from 0 to 3 and Latin indices from 1 to 3. 

These differential  equations (\ref{EME}--\ref{CE}) are supplemented 
by equations of state and the perfect conductivity condition

\begin{equation} 
             \bF_{\nu\mu} \bu^\mu = 0 . 
\label{UCC} 
\end{equation} 
This condition makes the second group of Maxwell equations redundant.
From (\ref{UCC}) one obtains 
\begin{equation} 
  \FF = 0 ,
\label{UCC1}
\end{equation} 

\begin{equation} 
  \bF\cdot\bF > 0 , 
\label{UCC2}
\end{equation} 
where $\bF\cdot\bF$ means  
$\bF_{\mu\nu}\bF^{\mu\nu}$.
Just like classical  MHD, RMHD is a hyperbolic  system of conservation
laws and has waves of four  types -- entropy, fast, Alfv\'en, and slow
(e.g Anile 1989; Komissarov 1999).

If the contribution  of
matter to the  total tensor of stress-energy momentum  is very small
then a perturbation technique suggests itself.
The zero order equations constitute the system of FFDE:

\begin{equation} 
  \nabla_\mu \Fs^{\mu\nu} = 0, 
\label{FE1}
\end{equation} 

\begin{equation} 
  \nabla_\mu \bT^{\nu\mu}_{(f)} = 0, 
\label{EME1}
\end{equation}
From (\ref{Temf},\ref{FE1}) one has 
\begin{equation} 
 \nabla_\mu \bT^{\nu\mu}_{(f)} = 
  -\bF_{\nu\mu}\bmath{J}^\mu ,
  \label{lor}
\end{equation} 
where 
\[ 
  \bmath{J}^\mu = \nabla_\nu \bF^{\mu\nu}  
\] 
and, thus, (\ref{EME1}) requires the vanishing of the Lorentz force.  
Since equations  (\ref{FE1},\ref{EME1}) do not involve $\bu_{\alpha}$, 
the perfect conductivity  condition should be replaced with conditions 
(\ref{UCC1},\ref{UCC2}), which express the degeneracy of electromagnetic 
field in FFDE. They ensure the existence of observers who detect  
only magnetic field.  
Let $\bmath{a}^\mu$ to be the velocity 
of such observer. It easy to see that  
\[ 
   \bF_{\mu\nu} \bmath{a}^{\nu} =0 
\]   
and, thus, $\bmath{a}^\nu$ is a zero eigenvector of 
$\bF_{\mu\nu}$. The second zero eigenvector can then 
be introduced via
\[ 
   \bmath{b}^\mu=\Fs^{\nu\mu} \bmath{a}_\nu. 
\]
In terms of $\bmath{a}^\mu$ and $\bmath{b}^\mu$ 

\begin{equation} 
  \bF_{\mu\nu} = \bmath{e}_{\mu\nu\alpha\beta} 
                       \bmath{a}^{\alpha}  \bmath{b}^\beta ,
\label{bu1}
\end{equation}  
and 
\begin{equation} 
  \Fs^{\mu\nu} = 2 \bmath{b}^{[\mu} \bmath{a}^{\nu]} . 
\label{bu2}
\end{equation}  
From the last equation it follows that the unit space-like vector 
$\bmath{c}^\mu$ orthogonal to $\bmath{a}^\mu$ and $\bmath{b}^\mu$ 
is a zero eigenvector of $\Fs_{\mu\nu}$. The second zero 
eigenvector, also space-like and orthogonal to $\bmath{a}^\mu$ 
and $\bmath{b}^\mu$, is given by 

\[ 
   \bmath{d}^\mu=\bF^{\nu\mu} \bmath{c}_\nu. 
\]
In terms of these vectors 
\begin{equation} 
  \Fs_{\mu\nu} = \bmath{e}_{\mu\nu\alpha\beta} 
                       \bmath{c}^{\alpha}  \bmath{d}^\beta ,
\label{cd1}
\end{equation}  
and 
\begin{equation} 
  \bF^{\mu\nu} = 2 \bmath{c}^{[\mu} \bmath{d}^{\nu]} . 
\label{cd2}
\end{equation}

From (\ref{UCC1}) it follows that
there  are only  five  independent components  of $\bF_{\mu\nu}$.  
In  components,   (\ref{FE1})  splits  into  three  evolution
equations  and  one  differential   constraint  on  the  initial  data
($\nabla_i B^i  = 0$). Thus, if  the system (\ref{UCC1})--(\ref{EME1})
is  self-consistent   then  (\ref{EME1})  has   only  two  independent
components.  Indeed, equations (\ref{lor},\ref{bu1}) show that  
that in the limit of degenerate electrodynamics, the vector 
$\nabla_\mu \bT^{\nu\mu}$ always belongs to 
the two-dimensional vector space generated by  
$\bmath{c}^\mu$ and $\bmath{d}^\mu$. 

The  evolution equations of the system  
(\ref{FE1},\ref{EME1}) have the form of conservation  laws.  In the
following section we shall show that this system is also hyperbolic.

\section{Hyperbolicity of Degenerate Electrodynamics} 

In   order   to    demonstrate   the   hyperbolicity   of  FFDE,
it  is   sufficient  to  study  its  one-dimensional
equations in  a locally pseudo-Cartesian system of coordinates.  In
such a frame, the electric and magnetic fields are defined via
\begin{equation} 
  E_i=F_{i0},\quad  B^i=\frac{1}{2}\epsilon^{ijk}F_{jk},  
\end{equation}  
where $\epsilon^{ijk}$ is the 3-dimensional Levi-Civita symbol. 
Then 
\[ 
  T^{00}=\frac{1}{2}(B^2+E^2), \quad 
  T^{i0}=\epsilon^{ijk}E_jB_k,
\] 
\[ 
  T^{ij}=-(E^iE^j+B^iB^j)+\frac{\delta^{ij}}{2}(E^2+B^2), 
\]
and conditions (\ref{UCC1},\ref{UCC2}) read 

\begin{equation} 
  \bmath{\vec{E}\cdot\vec{B}} = 0 ,
\label{UC1}
\end{equation} 

\begin{equation} 
  B^2-E^2 > 0 .
\label{UC2}
\end{equation} 
Notice that we use units such that the speed of light and $\pi$ do not
appear in the equations.

Like in MHD, the number  of evolution  equations and dependent variables for 
the one-dimensional system is reduced by one. 
Indeed, if $\partial/\partial x^2=\partial/\partial x^3 =0$ then $x^0$- ($t$-)
and the $x^1$-components of (\ref{FE1}) require
\[ 
   B_1 = const.   
\] 
Thus, the original system of 
FFDE is reduced to four independent evolution equations for four 
dependent variables, 
$B_2$, $B_3$, and two components of the electric field. The selection of 
these two components depends on the direction of magnetic field. For example, 
if $B_1=0$ then one of them must be $E_1$. As a result, we have to study 
three different sets of 1D equations to cover all possible cases. 
One can avoid such complications by 
relaxing the degeneracy condition (\ref{UC1}) and modifying the equations 
in a way such that this condition is satisfied automatically if it is  
satisfied by the initial data. For example, one can select all three space 
components 
of the energy-momentum equation and add the term $B_i\partial (B_i E^i)/ 
\partial t$ to obtain 
\begin{equation}
    F_{i\mu} \frac{\partial F^{\mu t}}{\partial t} + 
    F_{i\mu} \frac{\partial F^{\mu 1}}{\partial x^1} + 
    B_i \frac{\partial (B_i E^i)}{\partial t} =0 .
\label{augment} 
\end{equation} 
Together with the two remaining components of (\ref{FE1}) these 
equations constitute what we shell call the ``augmented one-dimensional
system of FFDE''. 
It has five independent equations for five components of $\bF_{\nu\mu}$. 
Obviously, solutions of this augmented system satisfying the constraint 
(\ref{UC1}) are also solutions of the original system of FFDE. Moreover, 
contraction of (\ref{augment}) with $B^i$ gives  
\[ 
  (B^i E_i) \frac{\partial E^1}{\partial x^1} + 
  B^2 \frac{\partial (B^iE_i)}{\partial t} = 0 
\] 
This result shows that if $B^iE_i =0$ at $t=0$ then all time 
derivatives of $B^iE_i$ at $t=0$  
also vanish and, therefore, $B^iE_i=0$ at $t>0$ as well.  
 
All characteristic waves of the 
FFDE must be present in the augmented system because of the way it has 
been constructed. It is also expected to 
have one additional nonphysical wave due to higher number of evolution equations. 
This extra wave can easily be identified because across this wave  
$E_iB^i$ does not have to stay constant.  

In vector form the augmented system reads 

\begin{equation} 
  \bmath{A}\frac{\partial \bU}{\partial t} + 
   \bmath{C} \frac{\partial \bU}{\partial x^1} = 0 , 
\label{case1}
\end{equation} 
where  
\begin{equation} 
    \bU = (B_2,B_3,E_1,E_2,E_3)^t , 
\label{a1}
\end{equation}
is the vector of dependent variables, 
\begin{equation}
 \bmath{A} = \left( \begin{array}{lllll} 
        1 & 0 & 0 & 0 & 0\\
        0 & 1 & 0 &  0 & 0\\ 
        B_1 E_2 & B_1 E_3 & B_1^2 & B_1 B_2 - B_3  & B_1 B_3+B_2 \\ 
        B_2 E_2 & B_2 E_3 & B_2 B_1+B_3 & B_2^2    & B_2 B_3-B_1 \\ 
        B_3 E_2 & B_3 E_3 & B_3 B_1-B_2 & B_2 B_3+B_1 & B_3^2  
        \end{array}\right) , 
\end{equation} 

\begin{equation}
\bmath{C} = \left( \begin{array}{lllll} 
        0      & 0      & 0 & 0 & -1 \\
        0      & 0      & 0 & 1 &  0 \\ 
        -B_2    & -B_3 & E_1 & 0 &  0 \\ 
         B_1 & 0 & E_2 & 0 & 0  \\ 
        0  & B_1 & E_3 & 0 & 0  \\ 
        \end{array}\right) . 
\end{equation} 

\noindent
Given (\ref{UC1},\ref{UC2}) the eigenvalue problem 

\begin{equation}
   (\bmath{A}-\mu\bmath{C})\cdot\bmath{r} = 0 
\label{eprob}
\end{equation} 
has the following solutions
 
\begin{equation} 
   \mu_f^\pm = \pm 1, 
\label{ewspeed}
\end{equation} 

\begin{equation} 
   \mu_a^\pm = \frac{B_3E_2-B_2E_3 \pm\sqrt{B_1^2(B^2-E^2)}}
               {B^2},  
\label{alfven1}
\end{equation} 

\begin{equation} 
   \mu_n = 0. 
\label{arti}
\end{equation} 

It is  easy to  verify that in the limit of FFDE the wave speeds of  
the fast waves of relativistic MHD \cite{Anile,Kom99} 
reduce to $\mu_f^\pm$  and 
the wave  speeds of Alfv\'en waves of relativistic MHD
reduce  to $\mu_a^\pm$. Hence, $\mu_n$ corresponds to 
the nonphysical wave of the augmented system. 

If $B_1\not=0$ then the eigenvectors of the fast modes are 

\begin{equation} 
  \bmath{r}_f^\pm= \left(-\eta_f,\ \zeta_f,\ 0,\ 
                  \mu_f^\pm\zeta_f,\ \mu_f^\pm \eta_f \right)^t, 
\label{eigenew1}
\end{equation}  
where 
\begin{equation} 
 \eta_f = E_3+\mu_f^\pm B_2, \quad \zeta_f = E_2-\mu_f^\pm B_3 .
\end{equation} 
It is shown in Sec.4 that these waves have the same properties as 
linearly polarised electromagnetic waves in vacuum. 

The eigenvectors of the Alfv\'en modes are
\begin{equation}
  \bmath{r}_a^\pm= \left(\zeta_a,\ \eta_a,\  
                -\frac{(\eta_a^2+\zeta_a^2)}{B_1},\  
                 \mu_a^\pm\eta_a,\ -\mu_a^\pm\zeta_a 
                   \right)^t, 
\label{eigenaw1}
\end{equation} 
where 
\begin{equation} 
 \eta_a = E_3+\mu_a^\pm B_2, \quad \zeta_a = E_2-\mu_a^\pm B_3,
\end{equation} 

In  some  applications  it  might  be useful  to  determine  the  left
eigenvectors  as well  as the  right eigenvectors.  Unfortunately, the
author was unable to find a concise form for the left eigenvectors and
they are therefore not presented here.

Like  MHD, the system  of FFDE is  not strictly
hyperbolic,  that  is it  allows  multiple  eigenvalues under  certain
conditions. Such  degenerate\footnote{In this paper three different kinds 
of degeneracy are encountered: 1) the degeneracy of electromagnetic 
field itself (see Sec.1), 2) the multiplicity of eigenvalues of the Jacobean 
matrix (Sec.3), and 3) the so-called linear degeneracy of hyperbolic 
waves (see Sec.4). } 
cases require  separate treatment because
the eigenvectors given by (\ref{eigenew1}) and (\ref{eigenaw1}) become
either singular or linearly dependent.

\begin{enumerate} 
\item Provided 
\begin{equation}
\bmath{\vec{E}}_t=-\bmath{\vec{i}}_1\bmath{\times\vec{B}}, 
\label{dcon1}
\end{equation}
where $\bmath{\vec{i}_1}$ is the unit vector along the $x_1$-axis and  
$\bmath{\vec{E}}_t$  is the tangential component of  electric field, one
has
\[
  \mu_a^+=\mu_f^+ = +1.  
\] 
and the corresponding  eigenvectors form the  two-dimensional vector space
generated by 
\begin{equation} 
      (0,\ 1,\ 0,\ 1,\ 0)^t \quad\mbox{and}\quad
      (-1,\ 0,\ 0,\ 0,\ 1)^t. 
\label{eigendegen1}
\end{equation} 
Obviously, vacuum electromagnetic 
waves propagating in the positive direction have the same eigenspace. 
Thus, in this limit the 
right Alfv\'en wave also behaves like a linearly polarised electromagnetic 
wave in vacuum.  
Similarly, if 
\begin{equation}
\bmath{\vec{E}}_t=\bmath{\vec{i}}_1\bmath{\times\vec{B}}, 
\label{dcon2}
\end{equation}
one has 
\[
  \mu_a^- = \mu_f^+ = -1 , 
\] 
and the corresponding  eigenvectors form the  two-dimensional vector space
generated by 
\begin{equation} 
      (0,\ 1,\ 0,\ -1,\ 0)^t \quad\mbox{and}\quad
      (1,\ 0,\ 0,\ 0,\ 1)^t. 
\label{eigendegen2}
\end{equation} 
This eigenspace is the same as that of vacuum electromagnetic waves
propagating  in  the negative  direction.  We  notice that  conditions
(\ref{dcon1})  and  (\ref{dcon2})  cannot be satisfied simultaneously 
unless $\vec{\bmath{E}}=0$ and $\vec{\bmath{B}}_t=0$. 

\item If $B_1=0$ then 
\begin{equation}
   \mu_a^+=\mu_a^- =\mu_a = \frac{E_2}{B^3}=-\frac{E_3}{B_2}  
\label{alfven2}
\end{equation} 
with a two-dimensional vector space of eigenvectors. One can use 
\begin{equation}
  \bmath{r}= (1,\ 0,\ \frac{-B_2(\mu_a^2-1)}{E_1},\ 0,\ -\mu_a)^t 
\label{eigendegen3}
\end{equation} 
and 
\begin{equation}
  \bmath{r}= (0,\ 1,\ \frac{-B_3(\mu_a^2-1)}{E_1},\ \mu_a,\ 0)^t 
\label{eigendegen4}
\end{equation} 
as the basis of the eigenspace if $E_1\not=0$ or 
\begin{equation}
  \bmath{r}= (0,\ 0,\ 1,\ 0,\ 0)^t 
\end{equation} 
and 
\begin{equation}
  \bmath{r}= (B_3,\ -B_2,\ 0,\ -\mu_a B_2,\ -\mu_a B_3)^t 
\end{equation} 
if $E_1=0$. 
\end{enumerate}

\section{Properties of Waves of Degenerate Electrodynamics}  

\subsection{Fast waves} 

From (\ref{eigenew1}) one obtains the following system of 
differential equations describing the fast simple wave: 
\begin{equation}
\nonumber
  \frac{dB_2}{-(E_3+\mu_f B_2)} = \frac{dB_3}{E_2-\mu_f B_3}= 
   \frac{dE_1}{0} =  
   \frac{dE_2}{\mu_f(E_2-\mu_f B_3)} =\frac{dE_3}{\mu_f(E_3+\mu_f B_2)},   
\label{d1} 
\end{equation}
where $\mu_f$ is either $\mu_f^+ = +1$ or $\mu_f^- = -1$.  
These are easily integrated to obtain 
\begin{equation} 
   E_1 = \mbox{const}. 
\label{d7} 
\end{equation} 

\begin{equation} 
  \eta_f=E_3+\mu_f B_2 = \mbox{const}, 
\end{equation} 

\begin{equation} 
  \zeta_f=E_2-\mu_f B_3 = \mbox{const}, 
\end{equation} 

\begin{equation} 
  E_iB^i = \mbox{const}, 
\end{equation} 

If we introduce 
the vector 
\[ 
   \bmath{\vec{t}}_f =(0,\zeta_f,\eta_f)^t 
\]
then 
\begin{equation} 
   d\bmath{\vec{E}}_t \,\|\, \bmath{\vec{t}}_f \quad\mbox{and}\quad 
   d\bmath{\vec{B}}_t \,\bot\, \bmath{\vec{t}}_f, 
\label{d8} 
\end{equation} 
where $\bmath{\vec{B}}_t$ and $\bmath{\vec{E}}_t$ are the tangential components 
of the fields. 
Thus, these are transverse waves and they have  
the same properties as linearly polarised electromagnetic waves in vacuum. 

\subsection{Alfv\'en waves}

From (\ref{eigenaw1}) one obtains the equations of an Alfv\'en 
simple wave:  
\begin{equation} 
  \frac{dB_2}{E_2-\mu_a B_3}=\frac{dB_3}{E_3+\mu_a B_2}= 
   -\frac{B_1 dE_1}{\eta_a^2+\zeta_a^2}= 
   \frac{dE_2}{\mu_a(E_3+\mu_aB_2)} = 
   -\frac{dE_3}{\mu_a(E_2-\mu_a B_3)},   
\label{e1} 
\end{equation} 
where $\mu_a$ is either $\mu_a^+$ or $\mu_a^-$. First, we show that 

\begin{equation}   
    B^2-E^2 = \mbox{const} 
\label{e2} 
\end{equation} 
and 

\begin{equation}   
  \mu_a =\mbox{const}.   
\label{e3} 
\end{equation} 
Since both these properties are  Lorentz invariant it is sufficient to
show that  they hold in one  particular frame. In this  case, the most
convenient frame  is the ``fluid'' frame,  where $\bmath{\vec{E}}=
\bmath{\vec{0}}$ and, hence, $dE^2=0$. Moreover, (\ref{e1}) ensures
\[ 
   d(B^2)=d(B_3E_2-B_2E_3)=0  
\]
and the results (\ref{e2},\ref{e3}) are manifest.  

Next, equations (\ref{e1}) can easily be integrated to obtain
\begin{equation} 
  \eta_a=E_3+\mu_aB_2  = \mbox{const}, 
\label{e4} 
\end{equation} 

\begin{equation} 
  \zeta_a=E_2-\mu_aB_3  = \mbox{const}, 
\label{e5} 
\end{equation} 

\begin{equation} 
  \zeta_a E_2 + \eta_a E_3  = \mbox{const}, 
\label{e6} 
\end{equation} 

\begin{equation} 
  E_iB^i  = \mbox{const}, 
\label{e7} 
\end{equation} 

Since the normal component of electric field varies the Alfv\'en waves
of FFDE are  not  quite  transverse.  This  is
similar to what has been  found for Alfv\'en waves in relativistic MHD
\cite{Kom97}.  
These  waves can still be described  as  linearly
polarised. If we introduce the vector

\[ 
   \bmath{\vec{t}}_a =(0,\zeta_a,\eta_a)^t 
\]
then 
\begin{equation} 
   d\bmath{\vec{E}}_t \,\bot\, \bmath{\vec{t}}_a \quad\mbox{and}\quad  
   d\bmath{\vec{B}}_t \,\|\, \bmath{\vec{t}}_a .
\label{e10} 
\end{equation} 
From this equation  and (\ref{d8}) one can see  that the polarisations
of  Alfv\'en  and  electromagnetic   waves  propagating  in  the  same
direction are not, in general, mutually orthogonal.

If $B_1=0$  then from (\ref{alfven1})  it follows that  Alfv\'en waves
propagate with the drift velocity
$$ 
   \mu_a = (\vec{E}\times\vec{B})_1/B^2, 
$$ 
and,  therefore, must  be  somewhat  similar to  the  contact wave  of
relativistic  MHD.  Their properties are most transparent in the wave frame, 
where $\mu=0$. From (\ref{alfven2})  one can see that in such frame
\begin{equation} 
    \bmath{\vec{E}}_t=\bmath{\vec{0}}.  
\end{equation} 
Using this result and (\ref{eigendegen3}) one finds that 
\begin{equation}
   (-B_1,\ B_2,\ 0,\ 0,\ 0)^t  \quad\mbox{and}\quad 
   (E_1,\ 0,\ B_2,\ 0,\ 0)^t  
\label{b3}
\end{equation} 
is a basis of the  Alfv\'en eigenspace. The first vector describes the
familiar rotation of $\bmath{\vec{B}}_t$.   The second allows variation of
the magnitude of $\bmath{\vec{B}}_t$, but requires
\[
  B_t^2-E_1^2 =\mbox{const} . 
\]

\begin{figure} 
\leavevmode 
\epsffile[0 0 256 256]{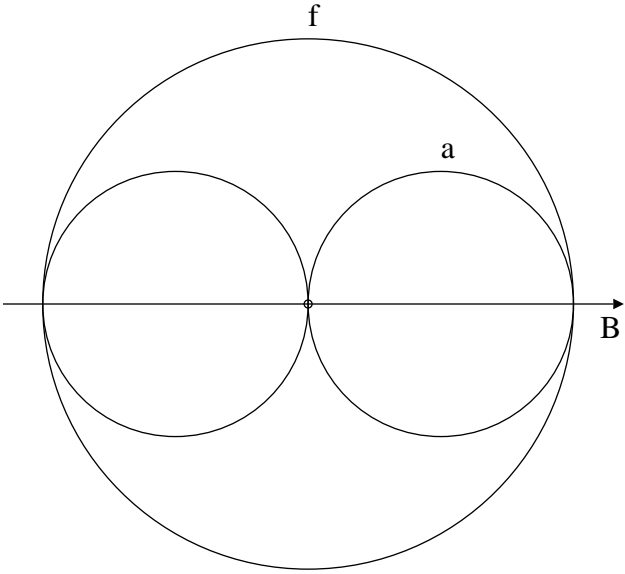} 
\caption{The normal speed diagram in the ``fluid'' frame 
($\bmath{\vec{E}}=\bmath{\vec{0}}$).}
\label{fig_nsp}
\end{figure}

Equations (\ref{ewspeed},\ref{e3}) show that all  waves of FFDE
are linearly degenerate, that is 
\[ 
   \bmath{r}\cdot\bmath{\nabla}_U \mu =0 ,
\]
and, therefore,  this system
does not allow  the formation of shocks via  wave steepening. 
However, discontinuities may be introduced via discontinuous initial data 
and/or boundary conditions.  For linearly degenerate waves the 
shock equations coincide with jump equations 
of simple waves \cite{Boillat,Anile}. 

Finally, figure 1 shows the normal wave speed diagram in the ``fluid'' 
frame, where $\vec{E}=\vec{0}$. Since in this frame 
\[ 
     \mu_a^\pm =\pm \cos\theta, 
\]
where $\theta$ is the angle between $\bmath{\vec{B}}$ and the wave vector, 
all curves on this diagram are circles.

\section{1D Upwind Numerical Scheme} 

Nowadays, the construction of  upwind numerical schemes for hyperbolic
conservation  laws is a well  established area  of numerical  analysis.
Here  we   describe a simple 
one-dimensional scheme which is very similar to the one constructed by 
Komissarov \shortcite{Kom99} for RMHD.  The space-time is considered 
to be flat.

\subsection{General Structure}

The one-dimensional  conservation laws of FFDE can be written in the 
form    

\begin{equation}
{{\partial \bQ} \over {\partial t}} + 
{{\partial \bP} \over {\partial x^1}} = 0, 
\label{f1}
\end{equation}

\noindent
where the vector of conserved variables is  

$$
\bQ =
   (S_1,S_2,S_3,B_2,B_3)^t,  
$$
where  $\bmath{\vec{S}}$ is the Poynting flux vector, 
and the vector of hyperbolic fluxes is 

$$
\bP =
  (T_{11},T_{12},T_{13},-E_3, E_2)^t . 
$$
Notice that  we use all three space  components of  the
energy-momentum equation. 
In addition, we introduce the auxiliary vector of ``primitive'' 
variables 
$$
 \bU=(B_2,B_3,E_1,E_2,E_3)^t . 
$$ 
These primitive variables can be converted into the conservative ones 
and vice versa via 

\begin{equation}
\bmath{\vec{S}}=\bmath{\vec{E}}\times\bmath{\vec{B}} , 
\label{f2}
\end{equation}
\begin{equation}
\bmath{\vec{E}} ={ 1 \over B^2} \bmath{\vec{S}}\times\bmath{\vec{B}} .
\label{f3}
\end{equation} 
The second equation automatically ensures the degeneracy condition 
(\ref{UC1}). Unfortunately, this does not mean that the second 
condition, (\ref{UC2}), will also be satisfied and extra care 
would have to be taken in cases where $B^2-E^2$ is close to zero.  
At the moment we do not have the means to ensure (\ref{UC1}).

Let us define a  regular Cartesian grid such  that the $i$th
cell  is  centered  at $x_i = ih$  and  occupies the  region  between
$x_{i-1/2}=(i  -1/2)h$ and $x_{i+1/2}=(i  + 1/2)h$,  where $h$  is the
mesh spacing.  Now suppose that we  know the solution at $t = t_n$ and
we  want to  calculate it  at a  later time  $t_{n+1}$.   We integrate
equations (\ref{f1})  over the $i$th cell and  from $t = t_n$  to $t =
t_{n+1}= t_n +\Delta t$ to get

\begin{equation}
    \bQ_{i,n+1} - \bQ_{i,n}  
      + {\Delta t \over h} 
        (\bP_{i+{1\over 2},n+{1 \over  2}}  - 
         \bP_{i-{1\over 2},n+{1 \over 2}}) = 0 .
\label{f4}
\end{equation}

\noindent
Here

\[
   \bQ_{i,n} = {1 \over h} 
   \int\limits_{x_{i-\frac{1}{2}}}^{x_{i+\frac{1}{2}}} 
   {\bQ(x,t_n)} dx
\]

\noindent
is the mean value of $\bQ$ in the $i$th cell at time $t_n$ and

\[ 
     \bP_{i+{1 \over 2},n+{1 \over 2}} = 
     {1 \over \Delta t}  \int\limits_{t_n}^{t_{n+1}} 
     \bP (x_{i+{1 \over 2}},t)dt ,
\]

\noindent
is  the  time  average  of   the  flux  at  the  cell  interface  with
$x~=~x_{i}+{1 \over 2}h$.

In a  first order  Godunov-type scheme (Godunov  1959), it  is assumed
that at the beginning of each time step the solution is uniform within
each cell. This implies initial discontinuities at the cell interfaces
and the  time averaged  fluxes can therefore  be found by  solving the
corresponding    Riemann   problems.    In    fact,   the    solution,
$\bU^*_{i+1/2}$,   of  the  problem   at  the   interface  with
$x=x_{i+1/2}$, does not  depend on time but only  on the initial left,
$\bU_{i,n}$, and right, $\bU_{i+1,n}$, states.  Then the
first order fluxes can be computed from

\[
    \bP_{i+{1\over 2},n+{1 \over 2}}  =  
    \bP(\bU^*_{i+{1\over 2}}) .  
\]

In  order to achieve  second order  accuracy, we  use the  first order
scheme  to obtain the  solution, $\bQ_{i,n  +{1 \over  2}}$ and
hence $\bU_{i,n+{1 \over 2}}$  at the half time, $t_{n+{1 \over
2}}=t_n + \Delta  t/2$. We then use this  to compute average gradients
of primitive variables in each cell as follows

\[
  {\left( \frac{ \partial \bU} {\partial   x} 
         \right)}_{i,n+{1 \over 2}} =  
   \frac{1}{h} {\rm av} \left( 
   \Delta \bU_{i,n+{1 \over 2}}, 
   \Delta \bU_{i+1,n+{1 \over 2}} \right),  
\] 
where 
\[
   \Delta \bU_{i,n+{1 \over 2}} =  \left( 
           \bU_{i,n+{1  \over  2}} - \bU_{i-1,n+{1 \over 2}}
       \right),   
\]

\noindent
and ${\rm  av}(a,b)$ is a non-linear averaging  function whose purpose
is to reduce  the scheme to first order in  space in the neighborhood
of  discontinuities. 
Here we will adopt the same averaging function as in \cite{Falle91}:

\[
  {\rm av}(a,b) =  \left\{
   \begin{array}{ll}
      \displaystyle{{{(a^2b + ab^2)} \over {(a^2 + b^2)}}} & 
      {\rm if}~ ab \geq 0, \mbox{ and } a^2 + b^2 \not= 0, \\
           & \\
       0 & {\rm if}~ ab < 0 \mbox{ or } a^2 +b^2=0 , 
   \end{array}  \right.
\]

These gradients  can now be used to  set up the left  and right states
for the second order Riemann problems

\[
   \bU^l_{i+1/2}  =  \bU_{i,n + {1 \over 2}}  + 
   {1 \over 2} h {\left( \frac{\partial \bU}{ \partial x} 
   \right)}_{i,n+{1 \over 2}},  
\]
\[
   \bU^r_{i+1/2}  =  \bU_{i+1,n + {1 \over 2}}  -  
   {1 \over 2} h {\left( \frac{\partial \bU}{ \partial x} 
   \right)}_{i+1,n+{1 \over 2}}.  
\]
The            solution,            $\bU^*_{i+1/2}            =
\bU(\bU^l_{i+1/2},\bU^r_{i+1/2})$,     to    this
Riemann problem allows us to compute the second order fluxes

\[
   \bP_{i+{1\over 2},n+{1 \over 2}}  = \bP
      (\bU^*_{i+{1\over 2}}),
\]

\noindent
which are used to advance the solution through the full time step from
$t=t_n$ to $t=t_{n+1}$ according to (\ref{f4}).

\subsection{Riemann Solvers} 

By a Riemann problem we  mean the initial value problem for (\ref{f1})
with the initial conditions
\begin{equation} 
   \bU(x,0) = \left\{ 
   \begin{array}{cc} 
      \bU^l & \mbox{for} \quad x \le 0, \\
      \bU^r & \mbox{for} \quad x > 0 .
   \end{array} 
   \right. 
 \label{f0} 
\end{equation} 
For  a  general hyperbolic  system  this  problem  has a  self-similar
solution   which  involves  only   shocks  and   centered  rarefactions
(e.g. Landau \& Lifshitz 1959, Jeffrey \& Taniuti 1964). Since   
all waves of FFDE are linearly degenerate, centered simple waves do not 
exist and, thus,  the solution involves only discontinuities.

\subsubsection{Linear Riemann Solver} 

Since all waves are linearly degenerate we have 

\begin{equation} 
    \bU_{r} = \bU_l + \sum_{i=1,4} 
       {\cal L}_i \bmath{r}_i, 
\label{ls1}
\end{equation} 
where $\bmath{r}_i$ is the eigenvector of the upstream 
state of ith wave. For fast waves these states are already known,  
they are the left and the right states of the Riemann problem.  
Upstream states of Alfv\'en waves are different and depend 
on the amplitudes of fast waves.  In our linear Riemann solver 
we ignore this difference.  Then (\ref{ls1}) becomes a linear 
system of five equations for four unknowns, ${\cal L}_i$. To handle it,  
we ignore one of the equations.       
The corresponding four components of the solution at $x^1=0$ are then 
found via  

\begin{equation} 
    \bU^* = 
      \bU_l + \sum_{\mu_i<0} {\cal L}_i \bmath{r}_i =
      \bU_r - \sum_{\mu_i>0} {\cal L}_i \bmath{r}_i .
\end{equation} 
The remaining component is determined using the degeneracy 
condition (\ref{UC1}).

\subsubsection{Exact Riemann Solver} 

In most Riemann problems originated from numerical simulations 
the difference between left and right states is rather small and
linear Riemann solvers are sufficiently accurate.  
However, for handling strong discontinuities one might require an  
exact solution of the Riemann problem. Thanks to the simplicity 
of FFDE, an exact Riemann solver can be constructed rather easily.  

First we consider the case where none of  the states of 
the Riemann problem has multiple eigenvalues. 
From the results of Sec.4 it follows that only fast waves 
change the value of  $p=B^2-E^2$ and only Alfv\'en waves change the 
value of $E_1$. This implies that the Riemann problem reduces to determining 
the values of $p$ and $E_1$ 
in the region bounded by Alfv\'en waves,
$\bar{p}$ and $\bar{E}_1$.  
Given the left state, the values of $E_1$ and $p$ in the right 
state, and  $\bar{p}$ and $\bar{E}_1$  
the corresponding right state $\bV_r(\bar{p},\bar{E}_1)$ can be found by 
successive application of the jump equations from left to right: 
\begin{equation} 
  \frac{[B_2]}{-\eta_f}=\frac{[B_3]}{\zeta_f}= \frac{[E_1]}{0} = 
   \frac{[E_2]}{\mu_f\zeta_f} =\frac{[E_3]}{\mu_f\eta_f} = 
   \frac{[p]}{-2\mu_f(\eta_f^2+\zeta_f^2)},   
\label{nrs1} 
\end{equation} 
for fast waves, and 
\begin{equation} 
  \frac{[B_2]}{\zeta_a}=\frac{[B_3]}{\eta_a}= 
   -\frac{B_1 [E_1]}{\eta_a^2+\zeta_a^2}= 
   \frac{[E_2]}{\mu_a\eta_a} = 
   -\frac{[E_3]}{\mu_a\zeta_a} = 
    \frac{[p]}{0}   
\label{nrs2} 
\end{equation} 
for Alfv\'en waves. This gives us the system of nonlinear  
equations 
\begin{equation}  
      \bV_r(\bar{p},\bar{E}_1) = \bU_r .
\label{nrs3} 
\end{equation} 
which can then be solved iteratively. Once $\bar{p}$ and $\bar{E}_1$ 
are found the resolved state is determined via (\ref{nrs1},\ref{nrs2}).  

In degenerate cases the problem is even easier. 
\begin{enumerate} 
\item   
If $\mu_f^+ = \mu_a^+$ in the right state of the Riemann problem, 
then the right fast and Alfv\'en waves merge into a single degenerate 
wave across which  
both $p$ and $E_1$ are constant 
Thus we immediately obtain $\bar{p}=p_r$, $\bar{E}_1=E_{1_r}$ and 
then determine the resolved state in the same way as in the non-degenerate 
case.   
Similarly, if $\mu_f^-=\mu_a^-$ in the left state then  
$\bar{p}=p_l$, $\bar{E}_1=E_{1_l}$. 
\item 
It is easy to verify that for 
any of the degenerate waves, 
the corresponding degeneracy condition holds on both sides of the wave.  
Thus, if $\mu_f^+$  equals to $\mu_a^+$ in the right state  
and $\mu_f^-$  equals to $\mu_a^-$ in the left state then in the 
resolved state conditions (\ref{dcon1}) and
(\ref{dcon2}) are satisfied simultaneously. This immediately gives 
$\vec{\bmath{E}} =0$ and $\vec{\bmath{B}}_t=0$ in the resolved state. 
\item   
If $B_1=0$ then the Alfv\'en waves merge into one contact-like wave 
across which both $p$ and $\mu_a$ are constant. This leads to the 
following two equations: 
\begin{equation}  
      p^*_l({\cal L}_{f_l}) = p^*_r({\cal L}_{f_r}), 
\label{nrs4} 
\end{equation} 
\begin{equation}  
      \mu^*_{a_l}({\cal L}_{f_l}) = \mu^*_{a_r}({\cal L}_{f_r}), 
\label{nrs5} 
\end{equation} 
where * indicates states adjacent to the contact wave. Simple 
manipulations reduce this system to a quadratic equation for 
${\cal L}_{e_l}$ (or ${\cal L}_{e_r}$). It appears that only one of 
its solutions satisfies the condition (\ref{UC2}).        
\end{enumerate}

\subsection{Test calculations} 

\begin{figure} 
\leavevmode 
\epsffile[0 0 520 256]{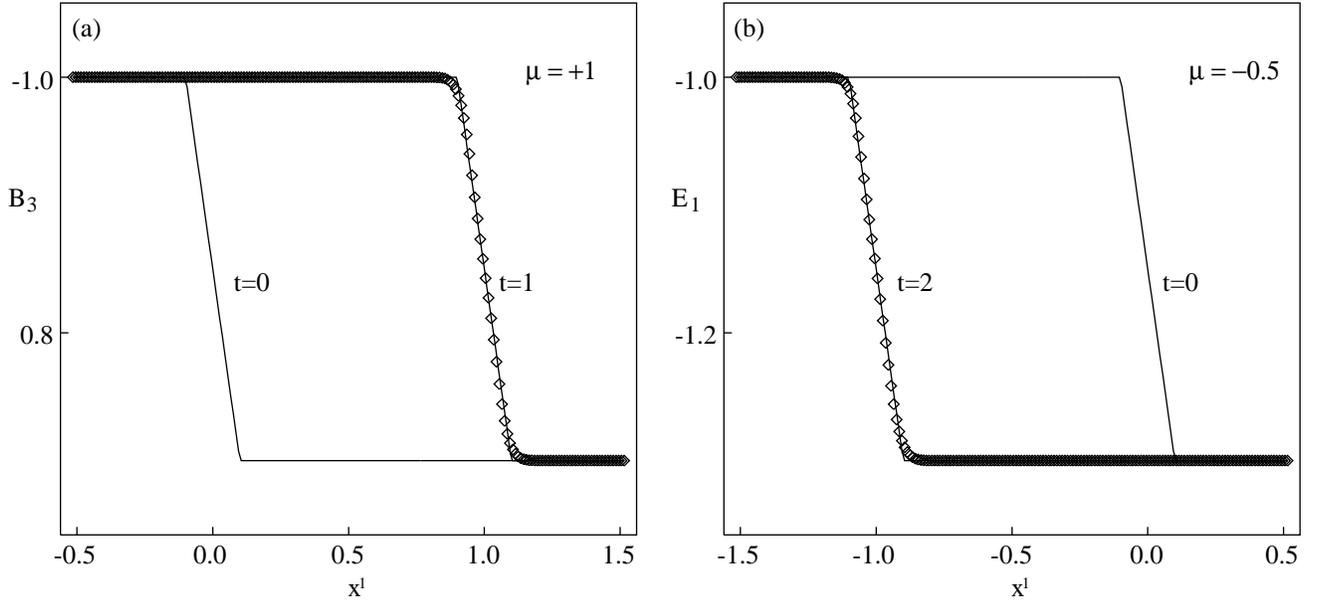} 
\caption{(a) The propagation of a fast wave. 
(b) The propagation of a non-degenerate Alfv\'en wave. 
The initial and the final exact solutions are shown by 
continuous lines and markers show the final numerical solution.}
\label{fig_ndw}
\end{figure}

\begin{figure} 
\leavevmode 
\epsffile[0 0 520 256]{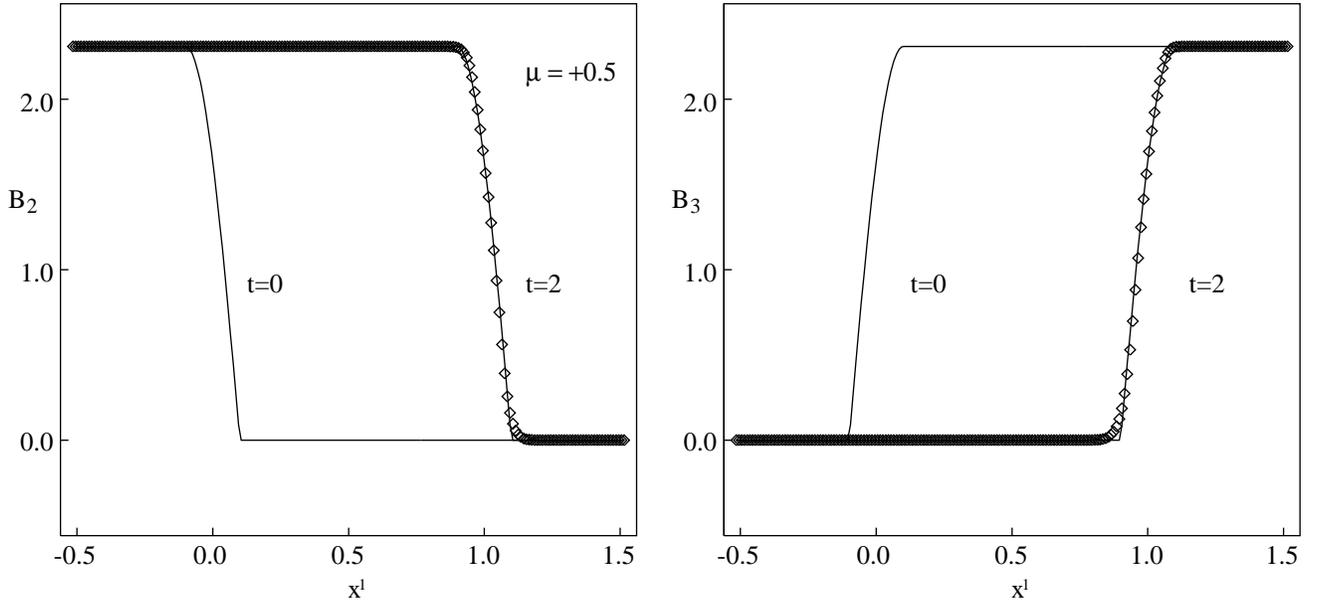} 
\caption{The propagation of a rotational degenerate Alfv\'en wave. 
The initial and the final exact solutions are shown by 
continuous lines and markers show the final numerical solution.}
\label{fig_daw}
\end{figure}

We have tested  the scheme against exact analytic  solutions for plane
electromagnetic  and Alfv\'en  waves including  all  degenerate cases.
The exact  solutions can  easily be constructed  using the  results of
Sec.3 and 4.   The profiles of these waves  should remain unchanged as
they  propagate.  Figures  \ref{fig_ndw}--\ref{fig_daw}  illustrate the
results  for some  of the  test  calculations.  In  these figures  the
initial and  the final exact  solutions are shown by  continuous lines
and markers show the final numerical solution.
 
\begin{description} 
\item{\it Fast wave:} \hskip 1cm
Figure 2a shows the results for a fast 
wave propagating in the positive direction. The initial solution is 
\[
 B_1=1.0, \quad B_3=E_2=0.0, 
\] 
\[ 
  B_2=\left\{\begin{array}{ll} 
       1.0 & \mbox{if}\quad x^1<-0.1 \\
       1.0+\frac{3}{2}(x^1+0.1) & \mbox{if}\quad -0.1<x^1<0.1 \\
       1.3 & \mbox{if}\quad x^1>0.1 \\
       \end{array} \right. 
\]
\item{\it Alfv\'en  wave:} Figure 2b shows the  results for a
non-degenerate  Alfv\'en wave  propagating in  the  negative direction
with the wave speed $\mu=-0.5$.  The initial solution is
\[ 
  B'_1=B'_2=1.0, \quad E'_2=E'_3=0.0 ,
\] 
\[ 
  B'_3=\left\{\begin{array}{ll} 
       1.0 & \mbox{if}\quad x^1<-0.1 \\
       1.0+\frac{3}{2}(x^1+0.1) & \mbox{if}\quad -0.1<x^1<0.1 \\
       1.3 & \mbox{if}\quad x^1>0.1 \\
       \end{array} \right.  ,
\]
where $\vec{B}'$ and $\vec{E}'$ are measured in the wave frame which is 
moving relative to the grid with speed $\mu=-0.5$. 

\item{\it    Degenerate   Alfv\'en   wave:}   
Figure (\ref{fig_daw}) shows the results for a rotational
degenerate Alfv\'en wave propagating to the right with $\mu=0.5$. 
The initial solution is
\[ 
   \vec{E}'=0, \quad B'_1=0, 
\] 
\[ 
    B'_2 =2\cos\phi, \quad B'_3 = 2\sin\phi, 
\] 
where 
\[ 
  \phi=\left\{\begin{array}{ll} 
       0.0 & \mbox{if}\quad x^1<-0.1 \\
       \frac{5\pi}{2}(x^1+0.1) & \mbox{if}\quad -0.1<x^1<0.1 \\
       \frac{\pi}{2} & \mbox{if}\quad x^1>0.1 \\
       \end{array} \right. .
\]

\item{\it Three waves problem:}  In this last test the initial discontinuity at 
$x=0$ splits into two fast discontinuities and a stationary Alfv\'en 
discontinuity.  The initial solution is  
\[ 
  \begin{array}{lll} 
       \vec{\bmath{B}}=(1.0, 1.5, 3.5) & 
       \vec{\bmath{E}}=(-1.0, -0.5, 0.5) & 
       \mbox{if}\quad x^1<0 ,\\
       \vec{\bmath{B}}=(1.0, 2.0, 2.(3)) & 
       \vec{\bmath{E}}=(-1.5, 1.(3), -0.5) & 
       \mbox{if}\quad x^1>0 . 
  \end{array}
\]

\end{description}

As one can see, numerical diffusion smoothes out the sharp wave fronts
of  the exact  solutions  but  otherwise the  agreement  is very  good
indeed. 
 
Not all Riemann problems with both the left and the right states 
satisfying the degeneracy conditions (\ref{UC1},\ref{UC2}) have solutions 
within the frame of degenerate electrodynamics. For example, the 
following problem does not seem have one   
\[ 
  \begin{array}{lll} 
       \vec{\bmath{B}}=(1.0, 1.0, 1.0) & 
       \vec{\bmath{E}}=(0.0, 0.5, -0.5) & 
       \mbox{if}\quad x^1<0 ,\\
       \vec{\bmath{B}}=(1.0, -1.0, -1.0) & 
       \vec{\bmath{E}}=(0.0, 0.5, -0.5) & 
       \mbox{if}\quad x^1>0 , 
  \end{array}
\] 
as iterations of our exact Riemann solver do not converge. 
To make sure that this is not due to a fault in the solver we have 
considered a related problem where 
this initial discontinuity is substituted with a linear transition layer. 
Numerical simulations (see figure \ref{fig_br}) show that within this 
layer $B^2-E^2$ decreases in 
time toward zero leading to the breakdown of the condition 
\[ 
   B^2-E^2 > 0.
\] 
As $B^2-E^2$ tends to zero the drift velocity of plasma 
\[ 
   \vec{\bmath{v}}_d = \vec{\bmath{E}}\times\vec{\bmath{B}}/B^2
\]
tends to the speed of light suggesting that one may no longer neglect 
the inertia of plasma particles.

\begin{figure} 
\leavevmode 
\epsffile[0 0 520 256]{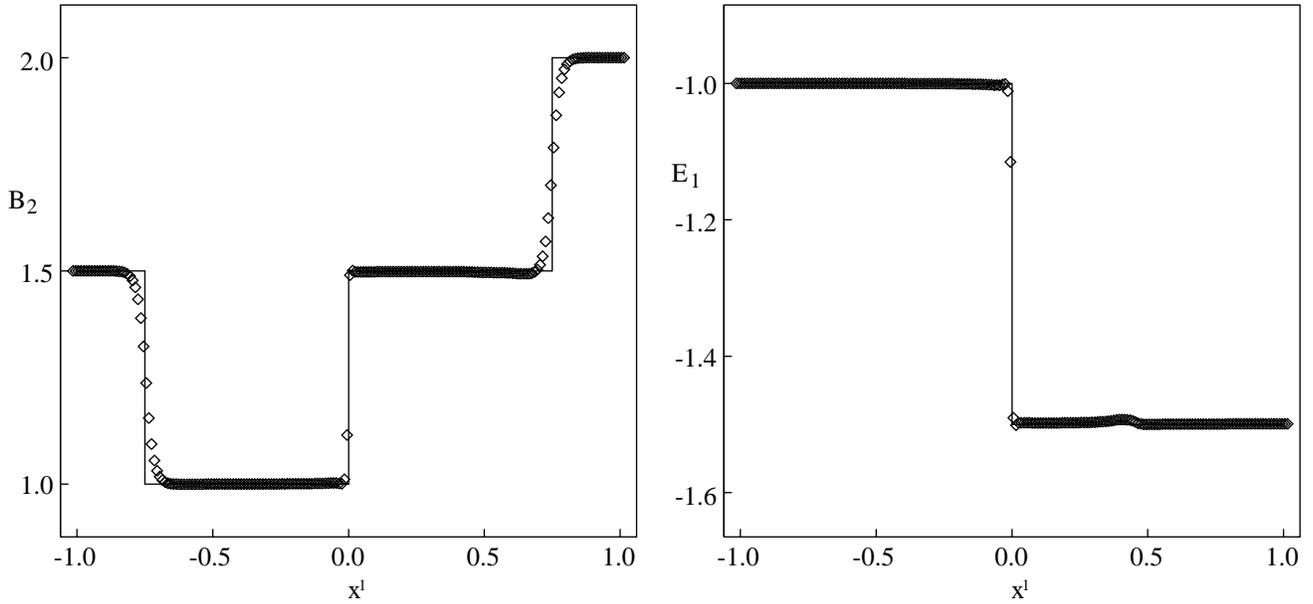} 
\caption{In this test problem the initial discontinuity at $x^1=0$ splits 
into two fast waves and one stationary Alfv\'en wave. 
The exact solution at $t=0.75$ is shown by 
continuous lines and markers show the corresponding numerical solution.}
\label{fig_sod}
\end{figure}

\begin{figure} 
\leavevmode 
\epsffile[0 0 256 256]{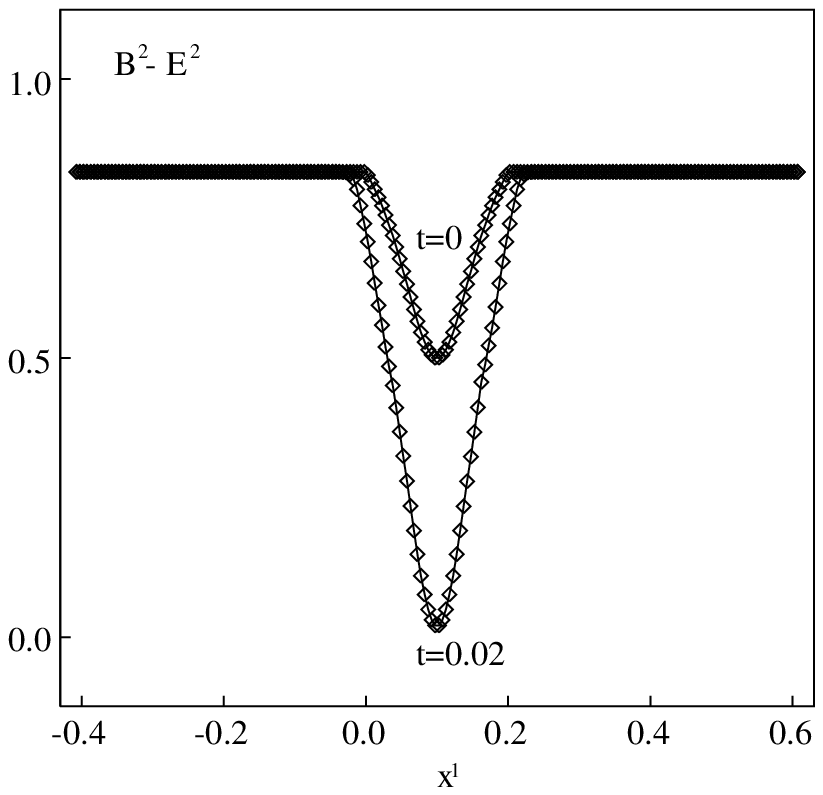} 
\caption{Breakdown of the approximation of degenerate electrodynamics as  
$B^2-E^2 \rightarrow 0$. See the last problem described in Sec.5.4 . }
\label{fig_br}
\end{figure}

\section{Conclusions} 

We   have   shown  that   the   evolution   equations  of  FFDE 
can be  written in the form of  conservation laws. One
group of equations arise  from the energy-momentum conservation. There
are only  two linear independent  dynamic equations (components of  one tensor
equation) in this  group.  This is in contrast to  MHD, where all four
components of the energy-momentum equation are linearly  independent, 
and vacuum electrodynamics, where
in general $\bF_{\mu\nu}$ does not have zero eigenvectors and the 
energy-momentum equation is reduced to 
\[ 
   \nabla_\nu F^{\mu\nu} = 0, 
\]   
which has three independent dynamic components.  
The one-dimensional system
of FFDE with plane symmetry, in flat space-time,
includes  four  linearly independent dynamic equations  for four  independent
components of the electromagnetic field tensor (other equations simply
require $B_1=$const.)  This system  is hyperbolic, though not strictly
hyperbolic, with  a pair  of fast waves 
and a pair  of Alfv\'en waves.  All characteristic  modes are linearly
degenerate, which means that this  system does not allow the formation
of shocks via steepening of continuous waves.

This formulation is particularly useful when it comes to developing 
numerical schemes for time-dependent FFDE as there has been a great deal 
of work on numerical methods for hyperbolic conservation 
laws. Here we have presented a simple one-dimensional scheme based 
mainly on our linear Riemann solver. Its generalization to 
multi-dimensions is relatively straightforward.    
We have  already constructed a 2D numerical
scheme adapted  to Kerr space-time which will be described elsewhere. 
Such schemes can be useful tools in studying  electromagnetically 
driven jets and winds from black holes and neutron stars and, perhaps, other 
related phenomena of relativistic astrophysics.  

\section{Acknowledgments} This research is partly supported by PPARC.
 The author thanks S.A.E.G.  Falle for useful discussions and comments
on the first version of this manuscript.

\end{document}